\documentclass[11pt]{article}

\def\mytitle#1{\setcounter{equation}{0}
\setcounter{footnote}{0}
\begin{flushleft}\Large\textbf{#1}\end{flushleft}
\vspace{0.25cm}}
\def\myname#1{\leftline{{\large #1}}\vspace{-0.13cm}}
\def\myplace#1#2{\small\begin{flushleft}\textit{#1}\\
\texttt{#2}\end{flushleft}}
\newenvironment{contribution}{\normalsize\noindent}{}
\def\myclassification#1{\small\noindent
Pacs no :
       #1\vspace{0.5cm}}
\usepackage{graphicx}
\begin{document}

\mytitle{Gravitational Collapse in Husain space-time for
Brans-Dicke Gravity Theory with Power-law Potential}

\vskip0.2cm \myname{Prabir Rudra\footnote{prudra.math@gmail.com}}
\myplace{Department of Mathematics, Pailan College of Management
and Technology, Bengal Pailan Park, Kolkata-700 104, India.}{}
\vskip0.2cm \myname{Ritabrata
Biswas\footnote{biswas.ritabrata@gmail.com}} \myplace{Department
of Mathematics, Indian Institute of Engineering Science and
Technology, Shibpur, Howrah-711 103, India.}{} \vskip0.2cm
\myname{Ujjal Debnath\footnote{ujjaldebnath@yahoo.com}}
\myplace{Department of Mathematics, Indian Institute of
Engineering Science and Technology, Shibpur, Howrah-711 103,
India.}{} \vskip0.2cm

\begin{abstract}
The motive of this work is to study gravitational collapse in
Husain space-time in Brans-Dicke gravity theory. Among many
scalar-tensor theories of gravity, Brans-Dicke is the simplest and
the impact of it can be regulated by two parameters associated
with it, namely, the Brans-Dicke parameter, $\omega$, and the
potential-scalar field dependency parameter $n$ respectively. V.
Husain's work on exact solution for null fluid collapse in 1996
has influenced many authors to follow his way to find the
end-state of the homogeneous/inhomogeneous dust cloud. Vaidya's
metric is used all over to follow the nature of future outgoing
radial null geodesics. Detecting whether the central singularity
is naked or wrapped by an event horizon, by the existence of
future directed radial null geodesic emitted in past from the
singularity is the basic objective. To point out the existence of
positive trajectory tangent solution, both particular parametric
cases(through tabular forms) and wide range contouring process
have been applied. Precisely, perfect fluid's EoS satisfies a wide
range of phenomena : from dust to exotic fluid like dark energy.
We have used the EoS parameter $k$ to determine the end state of
collapse in different cosmological era. Our main target is to
check low $\omega$ (more deviations from Einstein gravity-more
Brans Dicke effect) and negative $k$ zones. This particularly
throws light on the nature of the end-state of collapse in
accelerated expansion in Brans Dicke gravity. It is seen that for
positive values of EoS parameter $k$, the collapse results in a
black hole, whereas for negative values of $k$, naked singularity
is the only outcome. It is also to be noted that ``low $\omega$"
leads to the possibility of getting more naked singularities even
for a non-accelerating universe.
\end{abstract}
\myclassification{ 04.20.Dw(Singularities and Cosmic Censorship), 98.65.D(Dark energy)}

\section{Introduction}
Pioneered by the work of C. H. Brans, and R. H. Dicke (Brans, C.
H. et. al. 1961) in 1961 one thought (rather to say a metric
theory of gravitation) parallel to General Relativity (GR
hereafter) was importantly coined in literature. GR considers the
stress energy or matter tensor to be the source of the
gravitational field. The manner in which the pressure of
mass-energy acts in a region in case of Brans-Dicke theory, is
completely different from the way in which it acts in case of GR.
We know that in GR the geometric curvature of space-time totally
controls the motion of a body. But in case of Brans-Dicke theory
this dependence is somewhat alleviated. These salient features of
difference is testimony of the fact that Brans-Dicke theory is
quite different from the traditional theories of GR and as a
result triggers a lot of research. In Brans-Dicke theory an
additional scalar term $\phi$ is involved with the second ranked
tensor of energy mass. This $\phi$ introduces a physical effect
that changes the effective gravitational constant from place to
place. Brans-Dicke's construction was however supported by the
earlier works of Pascual Jordan (1959).\\

The field equations of Brans-Dicke theory contain a dimensionless
parameter $\omega$, known as Brans-Dicke coupling constant. This
is a tuneable parameter, the value of which can be adjusted to be
consistent with observational evidences. Brans-Dicke theory is
``less strignent" than GR in another sense: it admits more
solutions. In particular, exact vacuum solutions of the Einstein
field equation of GR, augmented by the trivial scalar field $\phi
=1$, become exact vacuum solutions in Brans-Dicke theory. But some
space times which are not vacuum solutions to the Einstein field
equation become solutions for Brans-Dicke theory, with appropriate
choice of scalar field, vacuum.\\

Like GR, Brans-Dicke theory predicts light deflection and the
precession of perihelia of planets orbiting the Sun. However, the
precise formulae which govern these effects, according to
Brans-Dicke theory, depend upon the value of the coupling constant
$\omega$. This means that it is possible to set an observational
lower bound on the possible values of $\omega$ from observations
of the solar system and other gravitational systems. It should be
stated that the value of $\omega$ consistent with experiment has
risen with time. In 1973, $\omega>5$ was consistent with known
data. By 1981, $\omega>30$ was consistent with known data. Viking
Space Probe says $\omega$ should exceed $500$ (from timing
experiments) (Reasenberg, R. D. et. al. 1979). In 2003 evidence
derived from the Cassini-Huygens experiment shows that the value
of $\omega$ must exceed $40,000$. It is often thought that GR is
obtained from the Brans-Dicke theory in the limit
$\omega\rightarrow \infty$ (Barrow, J. D. et. al. 1990). But
Faraoni claims that this breaks down when the trace of the
stress-energy momentum vanishes, i.e. $T_{\mu}^{\mu}=0$. Some have
argued that only general relativity
satisfies the strong equivalence principle.\\

Oppenheimer and Snyder, for the first time analyzed the collapse
of a dust cloud with a static Schwartzchild exterior and Friedmann
like interior (Oppenhiemer, J. R. et. al. 1939). In classical GR
gravitational collapse is a problem of great curiosity as we can
get at least two types of singularities from it. One, covered by
an event horizon, is coined as a black hole (BH hereafter) whereas
the singularity alone is popular as Naked Singularity (NS
hereafter). To determine the exact initial condition leading to
the formation of BH or NS is a thought experiment full of
challenge. Physically thinking the most important finding is a
physical initial condition leading down to the formation of a NS.
After all, one would always like to test the validity of cosmic
censorship hypothesis (CCH) laid down by R.Penrose, (Penrose, R.
1969) which stated that the end result of a collapse is bound to
be a singularity shrouded by an event horizon, i.e. a BH. Many
works (Eardley, D. M. et. al. 1979; Christodoulou, D. 1984;
Newman, R. P. A. C. 1986; Dwivedi, I. H. et. al. 1989; Waugh, B.
et. al. 1986; Ori, A. et. al. 1990) are there in last few decades
where the possibility of formation of NS has been investigated.
There is no general theory of the nature on the visibility of
singularities. Vaidya solution (Vaidya, P. C. 1951) is utilized on
many occasions to determine the end state of collapse. Harko et al
(Harko, T. et. al. 2000) have studied the gravitational collapse
of strange matter and analyzed the condition for formation of a NS
in the spherically symmetric Vaidya space-time. It has been shown
that whether a BH or a NS will be formed, is dependent upon many
issues like the initial distribution of density and velocity, the
constitutive nature of the collapsing matter etc. One of the
generalizations, among the many generalizations of Vaidya metric,
known as the Husain solution has been used to study the formation of a BH with short hair.\\

Maeda (Maeda, H. 2006) started the study of spherically symmetric
gravitational collapse without finding the explicit form of the
solution. S. Jhingan and S. G. Ghosh (Jhingan, S. et. al. 2010)
has shown that the different orders of curvature corrections can
cause sensible changes in the final fate of the gravitational
collapse (in the sense that massive NS is formed). This work was
followed by another quasi spherical gravitational collapse in 5D
Einstein Gauss Bonnet gravity (Ghosh, S. G. et. al. 2010).
Recently, another study of gravitational collapse in $f(R)$
gravity was studied by S. G. Ghosh and S. D. Maharaj (Ghosh, S. G.
et. al. 2012) where they have obtained a condition for the
occurrence of a NS in the collapse of null dust in higher
dimensional $f(R)$ gravity. T.P. Singh and P.S. Joshi in their
several papers for the first time used the technique of the
existence of outgoing null geodesic from the end state of collapse
to identify the central singularity is a naked singularity or a
singularity covered with an event horizon (Joshi, P. S. et. al.
1995; Singh, T. P. et. al. 1996; Joshi, P. S. et. al. 1992; Joshi,
P. S. et. al. 1993; Lake, K. 1992; Szekeres, P. et. al. 1993;
Joshi, P. S. 1993). Their method eventually became very popular
and several works on the same method was done in different gravity
theories and with different exotic fluids (Patil, K. D. et. al.
2005; Patil , K. D. et. al. 2006; Debnath, U. et. al. 2008;,
Debnath, U. et. al. 2004; Banerjee, A. et. al. 2003; Rudra,  P.
et. al. 2011; Debnath, U. et. al. 2012; Rudra, P. et. al. 2012).
Debnath et al in (Debnath, U. et. al. 2012) and Rudra et al in
(Rudra,  P. et. al. 2011) have studied the end fate of collapse in
higher dimensional gravity theory. They have shown that, farther
one goes from Einstein gravity, there is a greater chance of
having a NS. Here the effect of exotic fluid as the initial
substance of the collapsing cloud has been also studied and it has
been concluded that, more exotic the matter, more is the chance of
having NS. Scheel in (Scheel, M. A. et. al. 1995) demonstrated
that Openheimer-Snyder collapse in Brans-Dicke theory results in
BHs rather than NSs (which is true for a particular range of
$\omega$) with the positive values of $\omega$, they have
speculated that the apparent horizon of a BH can pass outside the
event horizon causing the decrease in the surface area over time.
The non negative values of $\omega$ forces the BH, to radiate its
scalar mass to infinity soon after the initial collapse. Otherwise
for negative $\omega$ the opposite incident occurs. In (Dong-il
Hwang et. al. 2010) we get another relevant and interesting work
regarding the gravitational collapse in the background of
Brans-Dicke theory of gravity discussing the effects
of different values of $\omega$.\\

Non-static spherically symmetric solutions of Einstein equations,
for a null fluid source was given by Husian (Husain, V. 1996) in
1996, where the density $\rho$ and pressure $p$ of the fluid is
related by $p=k\rho$. Two salient features of the solution were
that it is exact and inhomogeneous in nature. In Husain's work,
the $k > 1/2$ solutions describe the evolution of a naked
singularity into a black hole as the collapse proceeds. The
parameters in the Vaidya metric determine which of these
possibilities occur, and a black hole always forms at a finite
non-zero mass. The $k < 1/2$ solutions describe the collapse of
radiation from flat space to a black hole. All of the new
solutions supported the cosmic censorship conjecture. Later the
Vaidya solution was generalized by Wang et al (Wang, A. et. al.
1999). The solution included most of the known solutions of the
Einstein equation such as the anti-de-Sitter charged Vaidya
solution. Moreover the Husain solution has been extensively used
to study the formation of a black hole with short hair (Brown, J.
D. et. al. 1997). The most recent development in Husain solution
was witnessed when the gravitational collapse of the Husain
solution in four and five dimensional space-times was studied by
Patil et al (Patil , K. D. et. al. 2006).

Keeping all the previous works of gravitational collapse in GR/
different gravity theories in mind we feel it will be of a great
interest if we investigate the existence of radial null geodesics
from the collapsing body in the back ground of BD theory. The
scalar factor present in the theory may help in collapse to form a
NS more prominently than the GR does. Even we can use the BD
parameter $\omega$ as a regulator and can see what happens if we
deviate more and more from the Einstein gravity making $\omega$
sufficiently low. In this concern we must recall the fact that in
(Rudra, P. et. al. 2011) while working with Lovelock gravity, we
saw that greater the deviation from Einstein gravity greater was
the tendency to have the NS. So in this paper, we are mainly
studying the nature of singularities(BH or NS) formed by the
gravitational collapse in Brans-Dicke theory of gravity. In
section (\ref{calculation}), we present the brief overview of
generalized Husain solution in Brans-Dicke theory of gravity. We
will first construct the Einstein field equations in BD theory for
the Vaidya metric and then with a proper choice of the structural
dependence of the potential term upon the scalar field we will
determine the $m(t, r)$, the mass term. In the next two sections
we investigate the behaviour/existence of the outgoing radial null
geodesic from the singularity taking the Vaidya metric with the
mass term $m(t, r)$ derived in the last section. Finally, the
paper ends with some concluding remarks in section (\ref{Discussion}).\\

\section{Field Equations}\label{calculation}

The self-interacting Brans-Dicke theory is described by the
action: (choosing $8\pi G_{0}=c=1$) (Chakraborty, W. et. al. 2009)
\begin{equation}\label{Lag}
S=\int d^{4} x \sqrt{-g}\left[\phi R- \frac{\omega(\phi)}{\phi}
{\phi}^{,\alpha} {\phi,}_{\alpha}-V(\phi)+ {\cal L}_{m}\right]
\end{equation}
where $V(\phi)$ is the self-interacting potential for the BD
scalar field $\phi$ and the constant $\omega$ is the
BD parameter.

Here we consider the metric in  spherically symmetric space-time
in the form (Vaidya, P. C. 1951)
\begin{equation}\label{collapse2.1}
ds^{2}=-\left(1-\frac{m(t,r)}{r}\right)dt^{2}+2dtdr+r^{2}d\Omega_{2}^{2}
\end{equation}
where $r$ is the radial co-ordinate and $t$ is the null
co-ordinate, $m(t,~r)$ gives the gravitational mass inside the
sphere of radius $r$ and $d\Omega_{2}^{2}$ is the line element on
a unit 2-sphere.

From the Lagrangian density (\ref{Lag}) we obtain the field
equations (Chakraborty, W. et. al. 2009)
\begin{equation}
G_{\mu \nu}=\frac{\omega(\phi)}{{\phi}^{2}}\left[\phi  _{ , \mu}
\phi _{, \nu} - \frac{1}{2}g_{\mu \nu} \phi _{, \alpha} \phi ^{ ,
\alpha} \right] +\frac{1}{\phi}\left[\phi  _{, \mu ; \nu} -g_{\mu
\nu}~ ^{\fbox{}}~ \phi \right]-\frac{V(\phi)}{2 \phi} g_{\mu
\nu}+\frac{1}{\phi}T_{\mu \nu}
\end{equation}
and
\begin{equation}\label{tensor?!}
^{\fbox{}}~\phi=\frac{1}{3+2\omega}T-\frac{1}{3+2\omega}\left[2V(\phi)-\phi
 \frac{dV(\phi)}{d\phi}\right]
 \end{equation}
where $T=T_{\mu \mu}g^{\mu \mu}$.

Now we consider two types of fluids like Vaidya null radiation and
a perfect fluid having the form of the energy momentum tensor
\begin{equation}\label{collapse2.3}
T_{\mu\nu}=T_{\mu\nu}^{(n)}+T_{\mu\nu}^{(m)}
\end{equation}
with
\begin{equation}\label{collapse2.4}
T_{\mu\nu}^{(n)}=\sigma l_{\mu}l_{\nu}
\end{equation}
and
\begin{equation}\label{collapse2.5}
T_{\mu\nu}^{(m)}=(\rho+p)(l_{\mu}\eta_{\nu}+l_{\nu}\eta_{\mu})+pg_{\mu\nu}
\end{equation}

Where, $\rho$ and $p$ are the energy density and pressure for the
perfect fluid and $\sigma$ is the energy density corresponding to
Vaidya null radiation. In the co-moving co-ordinates
($v,r,\theta_{1},\theta_{2},...,\theta_{n}$), the two eigen
vectors of energy-momentum tensor namely $l_{\mu}$ and
$\eta_{\mu}$ are linearly independent future pointing null vectors
having components
\begin{equation}\label{collapse2.6}
l_{\mu}=(1,0,0,...,0)~~~~ and~~~~
\eta_{\mu}=\left(\frac{1}{2}\left(1-\frac{m}{r^{n-1}}\right),-1,0,...,0
\right)
\end{equation}
and they satisfy the relations
\begin{equation}\label{collapse2.7}
l_{\lambda}l^{\lambda}=\eta_{\lambda}\eta^{\lambda}=0,~
l_{\lambda}\eta^{\lambda}=-1
\end{equation}

\section{The Solution}

The Einstein field equations ($G_{\mu\nu}=T_{\mu\nu}$) for the
metric (\ref{collapse2.1}) and the wave equation for the BD scalar
field $\phi$ are the following

$$\frac{\left(r-m\right)m'+r\dot{m}}{r^{3}}=\frac{\omega}{\phi^{2}}\left[\dot{\phi}^{2}+\frac{1}{2}\left(1-\frac{m}{r}\right)\left\{2\dot{\phi}\phi'+\phi'^{2}\left(1-\frac{m}{r}\right)\right\}\right]+\frac{1}{\phi}\left[\ddot{\phi}-\frac{\dot{\phi}}{2r}\left(\frac{m}{r}-m'\right)\right.$$

$$\left.-\frac{\phi'}{2r}\left(\frac{m}{r}-\frac{m^{2}}{r^{2}}-m'+\frac{mm'}{r}+\dot{m}\right)+\left(1-\frac{m}{r}\right)\left\{2(\dot{\phi})'+\phi''\left(1-\frac{m}{r}\right)-\frac{\phi'}{r}\left(2-\frac{3m}{r}+m'\right)\right\}\right]$$
\begin{equation}\label{1}
+\frac{V(\phi)}{2\phi}\left(1-\frac{m}{r}\right)+\frac{1}{\phi}\left[\sigma+\rho\left(1-\frac{m}{r}\right)\right]~,
\end{equation}

\begin{equation}\label{2}
\frac{\omega \phi'^{2}}{\phi^{2}}+\frac{\phi''}{\phi}=0,
\end{equation}

$$-\frac{1}{2}rm''=\frac{\omega}{\phi^{2}}\left[-\frac{r^{2}}{2}\phi'\left\{2\dot{\phi}+\phi'\left(1-\frac{m}{r}\right)\right\}\right]+\frac{1}{\phi}\left[r\dot{\phi}-m\phi'-2r^{2}(\dot{\phi})'-r^{2}\phi''\left(1-\frac{m}{r}\right)+r\phi'\left(3-3\frac{m}{r}+m'\right)\right]$$
\begin{equation}\label{3}
-\frac{V(\phi)}{2\phi}r^{2}+\frac{pr^{2}}{\phi}~,
\end{equation}

\begin{equation}\label{4}
-\frac{m'}{r^{2}}=\frac{\omega}{\phi^{2}}\left[-\frac{\phi'^{2}}{2}\left(1-\frac{m}{r}\right)\right]+\frac{1}{\phi}\left[-(\dot{\phi})'+\frac{\phi'}{r}\left(\frac{m'}{2}-\frac{5m}{2r}+2\right)-\phi''\left(1-\frac{m}{r}\right)-\frac{V(\phi)}{2}-\rho\right]
\end{equation}
 and

\begin{equation}\label{5}
2(\dot{\phi})'+\phi''\left(1-\frac{m}{r}\right)-\phi'\left(\frac{2}{r}-\frac{3m}{r^{2}}+\frac{m'}{r}\right)=\frac{\rho-3p}{3+2\omega}+\frac{1}{3+2\omega}\left[2V(\phi)-\phi\frac{dV(\phi)}{d\phi}\right]
\end{equation}
\vspace{5mm} where an over-dot and dash stand for differentiation
with respect to $t$ and $r$ respectively.

{\bf Here we use the power law form of potential in the Brans-Dicke
theory as given below \textbf{(Bisabr, Y. 2012, Chattopadhyay, S.
2013, Yang, W-Q. et al. 2011)}. While studying the evolution of naked singularities in Brans-Dicke cosmology,  Ziaie, A. H. et al (2010) has shown that for particular matter fields if $\phi=a^\alpha$ then $V(\phi)$ takes the form $\beta \phi^{-\frac{3(1+\omega_{BD})}{\alpha}}$ where $\beta=2+\frac{\alpha}{3}(1+\rho_{0m})(6+\omega_{BD})$, here $\rho_{0m}$ stands for present time barotropic mass density. Ultimately we can generalise the field dependency of the potential as }
\begin{equation}\label{6}
V(\phi)=V_{0}\phi^{n}
\end{equation}
On solving field equation (\ref{2}), the expression for $\phi$ is
obtained as
\begin{equation}\label{7}
\phi=B(t)r^{\frac{1}{\omega+1}}
\end{equation}

where $B(t)$ is the arbitrary function of $t$. We assume the
matter fluid obeys the barotropic equation of state
\begin{equation}\label{8}
p=k\rho,~~~(k,~a~constant)
\end{equation}

Using  equations (\ref{3}), (\ref{4}), (\ref{6}) and (\ref{8}), we
have the differential equation in $m$ as

$$\left(\frac{1}{2kr}\right)m''+\left[\frac{k+2+2k\left(\omega+1\right)}{2r^{2}k\left(\omega+1\right)}\right]m'-\left[\frac{6k\omega+5k+9\omega+8}{2k\left(\omega+1\right)^{2}r^{3}}\right]m$$
\begin{equation}\label{9}
+\left[\frac{5\omega
k+4k+7\omega+6}{2kr^{2}\left(\omega+1\right)^{2}}-\frac{\dot{B}\left(1+k\right)}{Brk\left(\omega+1\right)}-\frac{V_{0}}{2k}\left(1+k\right)\phi^{n-1}\right]=0
\end{equation}
Solving the above differential equation we obtain the explicit
solution for $m$ as,
$$m(t,r)=f_{1}(t)r^{\omega_{1}}+f_{2}(t)r^{\omega_{2}}+\frac{5\omega k+4k+7\omega+6}{\left(1-\omega_{1}\right)\left(1-\omega_{2}\right)}r-\frac{2\dot{B}\left(1+k\right)\left(1+\omega\right)}{B\left(2-\omega_{1}\right)\left(2-\omega_{2}\right)}r^{2}$$
\begin{equation}\label{10}
-\frac{V_{0}B^{n-1}\left(\omega+1\right)^{4}\left(k+1\right)}{\left\{\left(\omega+n\right)-\omega_{1}\left(\omega+1\right)\right\}\left\{\left(\omega+n\right)-\omega_{2}\left(\omega+1\right)\right\}}r^{\frac{\omega+n}{\omega+1}}
\end{equation}
where
\begin{equation}\label{11}
\omega_{1},
\omega_{2}=\frac{\left(\omega-3k-2k\omega-1\right)\pm\sqrt{\left\{\left(k+2\right)+\left(2k-1\right)\left(\omega+1\right)\right\}^{2}-4\left(6k\omega+5k+9\omega+8\right)}}{2\left(\omega+1\right)}
\end{equation}
Here $f_{1}(t)$ and $f_{2}(t)$ are arbitrary functions of $t$.

Therefore the metric (\ref{collapse2.1}) can be written as
$$ds^{2}=\left[1-f_{1}(t)r^{\omega_{1}-1}-f_{2}(t)r^{\omega_{2}-1}-\frac{5\omega k+4k+7\omega+6}{\left(1-\omega_{1}\right)\left(1-\omega_{2}\right)}+\frac{2\dot{B}\left(1+k\right)\left(1+\omega\right)}{B\left(2-\omega_{1}\right)\left(2-\omega_{2}\right)}r\right.$$
\begin{equation}\label{12}
\left.+\frac{V_{0}B^{n-1}\left(\omega+1\right)^{4}\left(k+1\right)}{\left\{\left(\omega+n\right)-\omega_{1}\left(\omega+1\right)\right\}\left\{\left(\omega+n\right)-\omega_{2}\left(\omega+1\right)\right\}}r^{\frac{n-1}{\omega+1}}\right]dt^{2}+2dtdr+r^{2}d\Omega_{2}^{2}
\end{equation}

which is called the the Husain metric or Generalized Vaidya metric
in Brans-Dicke gravity.

\section{Collapse Study}
We shall discuss the existence of NS in generalized Vaidya
space-time by studying radial null geodesics. In fact, we shall
examine whether it is possible to have outgoing radial null
geodesics which were terminated in the past at the central
singularity $r=0$. The nature of the singularity (NS or BH) can be
characterized by the existence of radial null geodesics emerging
from the singularity. The singularity is at least locally naked if
there exist such geodesics and if no such geodesics exist it is a
BH.

A singularity caused by a catastrophic gravitational collapse may
be a NS or a BH. Famous Hawking-Penrose singularity theorems
provide no information on that issue. The cosmic censorship
hypothesis essentially states that , in GR, the end state of
gravitational collapse is always a BH : the gravitational
singularity must necessarily be covered by an event horizon.
However, it comes with no known proof. Inhomogeneous dust cloud
may result in a naked singularity through a collapse (Eardley, D.
M. et. al. 1979). Fluids with different EoS other than dust also
give rise to interesting results (Joshi, P. S. et. al. 1992).
Keeping the above literature in view the censorship hypothesis
needs to get generalised (Joshi, P. S. et. al. 1995).

Let $R(t, ~r )$ is the physical radius at time $t$ of the shell
labelled by $r$. Scaling freedom says at the starting epoch $t=0$
we should have $R(0,~r)=r$. We should keep it in mind that in the
inhomogeneous case (more generalized one), different shells become
singular at different times. Now if there are future directed
radial null geodesics coming out of the singularity , with a well
defined tangent at  the singularity $\frac{dR}{dr}$ must tend to a
finite limit in the limit of approach to the singularity in the
past along these trajectories.

The points $(t_0, ~r)=0$ occurs, where the singularity $R(t_0,
0)=0$ occurs corresponds to the physical situation where matter
shells are crushed to zero radius. This kind of singularity
($r=0$) is known to be a central singularity.

The singularity is an NS if there are future directed non-space
like curves in the space time with their past end points at the
singularity. Now if the outgoing null geodesics are to terminate
in the past at the central singularity at $r=0$ at $t=t_0$ where
$R(t_0, 0)=0$, then along these geodesics we should have (Singh,
T. P. et. al. 1996) $R\rightarrow 0$ as $r\rightarrow 0$.

The equation for outgoing radial null geodesics can be obtained
from equation (\ref{collapse2.1}) by putting $ds^{2}=0$ and
$d\Omega_{2}^{2}=0$ as
\begin{equation}\label{collapse2.24}
\frac{dt}{dr}=\frac{2}{\left(1-\frac{m(t,r)}{r}\right)}.
\end{equation}
It can be seen easily that $r=0,~t=0$ corresponds to a singularity
of the above differential equation. Suppose $X=\frac{t}{r}$ then
we shall study the limiting behavior of the function $X$ as we
approach the singularity at $r=0,~t=0$ along the radial null
geodesic. If we denote the limiting value by $X_{0}$ then
\begin{eqnarray}\label{collapse2.25}
\begin{array}{c}
X_{0}\\\\
{}
\end{array}
\begin{array}{c}
=lim~~ X \\
\begin{tiny}t\rightarrow 0\end{tiny}\\
\begin{tiny}r\rightarrow 0\end{tiny}
\end{array}
\begin{array}{c}
=lim~~ \frac{t}{r} \\
\begin{tiny}t\rightarrow 0\end{tiny}\\
\begin{tiny}r\rightarrow 0\end{tiny}
\end{array}
\begin{array}{c}
=lim~~ \frac{dt}{dr} \\
\begin{tiny}t\rightarrow 0\end{tiny}\\
\begin{tiny}r\rightarrow 0\end{tiny}
\end{array}
\begin{array}{c}
=lim~~ \frac{2}{\left(1-\frac{m(t,r)}{r}\right)} \\
\begin{tiny}t\rightarrow 0\end{tiny}~~~~~~~~~~~~\\
\begin{tiny}r\rightarrow 0\end{tiny}~~~~~~~~~~~~
 {}
\end{array}
\end{eqnarray}
Using equations (\ref{10}) and (\ref{collapse2.25}), we have
\begin{eqnarray*}
\frac{2}{X_{0}}=
\begin{array}llim\\
\begin{tiny}t\rightarrow 0\end{tiny}\\
\begin{tiny}r\rightarrow 0\end{tiny}
\end{array}\left[1-f_{1}(t)r^{\omega_{1}-1}-f_{2}(t)r^{\omega_{2}-1}-\frac{5\omega
k+4k+7\omega+6}{\left(1-\omega_{1}\right)\left(1-\omega_{2}\right)}+\frac{2\dot{B}\left(1+k\right)\left(1+\omega\right)}{B\left(2-\omega_{1}\right)\left(2-\omega_{2}\right)}r\right.
\end{eqnarray*}
\begin{equation}
\left.+\frac{V_{0}B^{n-1}\left(\omega+1\right)^{4}\left(k+1\right)}{\left\{\left(\omega+n\right)-\omega_{1}\left(\omega+1\right)\right\}\left\{\left(\omega+n\right)-\omega_{2}\left(\omega+1\right)\right\}}r^{\frac{n-1}{\omega+1}}\right]
\end{equation}
Now choosing $f_{1}(t)=\lambda
t^{-(\omega_{1}-1)}$,~~~$f_{2}(t)=\gamma t^{-(\omega_{2}-1)}$~~
and ~$B(t)=B_{0}t^{-\frac{1}{1+\omega}}$, we obtain the algebraic
equation of $X_{0}$ as
$$\frac{V_{0}\left(\omega+1\right)^{4}\left(k+1\right)B_{0}^{n-1}}{\left[\left(\omega+n\right)-\omega_{1}\left(\omega+1\right)\right]\left[\left(\omega+n\right)-\omega_{2}\left(\omega+1\right)\right]}X_{0}^{\frac{\omega-n+2}{\omega+1}}-\lambda
X_{0}^{2-\omega_{1}}-\gamma X_{0}^{2-\omega_{2}}$$
\begin{equation}
+\left[1-\frac{5\omega
k+4k+7\omega+6}{\left(1-\omega_{1}\right)\left(1-\omega_{2}\right)}\right]X_{0}-2\left[1+\frac{\left(1+k\right)}{\left(2-\omega_{1}\right)\left(2-\omega_{2}\right)}\right]=0
\end{equation}
Now if we get only non-positive solution of the equation we can
assure the formation of a BH. Getting a positive root indicates a
chance to get a NS. Since the obtained equation is a highly
complicated one, it is extremely difficult to find out an analytic
solution of $X_{0}$ in terms of the variables involved. So our
idea is to find out different numerical solutions of $X_{0}$, by
assigning particular numerical values to the associated variables.

The different solutions of $X_{0}$ for different sets of
parametric values ($\lambda,\gamma,V_{0},B_{0},n,\omega,k$) are
given here in a tabular form (Table 1a-e).\\

\begin{center}
Table1a
\begin{tabular}{|l|}
\hline\hline\\ ~~~~~~~~~~~~~~~~~~~~~~~~~~~~~~~~~~~~~~~~~~~~~~~~For $k=1$ (stiff perfect fluid)\\
\hline\hline
~~$\omega$~~~~~~~~~$\lambda$~~~~~~~~~$\gamma$~~~~~~~~~$V_{0}$~~~~~~~~~$B_{0}$~~~~~~~~~$n$
~~~~~~~~~~~~~~~Positive roots ($X_{0}$)
\\ \hline
\\
~~2~~~~~~~~~~1~~~~~~~~~~1~~~~~~~~5~~~~~~~~~~~~6~~~~~~~~~~~~1~~~~~~~~~~~~~~~~~~~~~~~~$-$
\\
~~,,~~~~~~~~~,,~~~~~~~~~,,~~~~~~~~,,~~~~~~~~~~~~,,~~~~~~~~~~~~~~2~~~~~~~~~~~~~~~~~~~~~~~~$-$
\\
~~,,~~~~~~~~~,,~~~~~~~~~,,~~~~~~~~,,~~~~~~~~~~~~,,~~~~~~~~~~~~~~4~~~~~~~~~~~~~~~~~~~~~~~~$-$
\\ \hline \\
~~1~~~~~~~~~0.1~~~~~~~~0.1~~~~~~~1~~~~~~~~~~~~1~~~~~~~~~~~~1~~~~~~~~~~~~~~~~~~~~~~~~$-$
\\
~~,,~~~~~~~~~,,~~~~~~~~~,,~~~~~~~~,,~~~~~~~~~~~~,,~~~~~~~~~~~~~~2~~~~~~~~~~~~~~~~~~~~~~~~$-$
\\
~~,,~~~~~~~~~,,~~~~~~~~~,,~~~~~~~~,,~~~~~~~~~~~~,,~~~~~~~~~~~~~~4~~~~~~~~~~~~~~~~~~~~~~~~$-$
\\ \hline \\
~~-2~~~~~~~~0.1~~~~~~~~0.1~~~~~~~1~~~~~~~~~~~~1~~~~~~~~~~~~1~~~~~~~~~~~~~~~~~~~~~~~~$-$
\\
~~,,~~~~~~~~~,,~~~~~~~~~,,~~~~~~~~,,~~~~~~~~~~~~,,~~~~~~~~~~~~~~2~~~~~~~~~~~~~~~~~~~~~~
~~$0.9449257$
\\
~~,,~~~~~~~~~,,~~~~~~~~~,,~~~~~~~~,,~~~~~~~~~~~~,,~~~~~~~~~~~~~~4~~~~~~~~~~~~~~~~~~~~~~~~$0.8592578$
\\ \hline \\
~~-3~~~~~~~~0.1~~~~~~~~0.1~~~~~~~1~~~~~~~~~~~~1~~~~~~~~~~~~1~~~~~~~~~~~~~~~~~~~~~~~~$-$
\\
~~,,~~~~~~~~~,,~~~~~~~~~,,~~~~~~~~,,~~~~~~~~~~~~,,~~~~~~~~~~~~~~2~~~~~~~~~~~~~~~~~~~~~~
~~$0.4381$
\\
~~,,~~~~~~~~~,,~~~~~~~~~,,~~~~~~~~,,~~~~~~~~~~~~,,~~~~~~~~~~~~~~4~~~~~~~~~~~~~~~~~~~~~~~~
$0.39825$
\\
 \hline\hline
 \end{tabular}

\end{center}
\begin{center}
Table1b
\begin{tabular}{|l|}

\hline\hline
\\ ~~~~~~~~~~~~~~~~~~~~~~~~~~~~~~~~~~~~~~~~~~~~~~~~For $k=1/3$ (radiation)\\
\hline\hline
~~$\omega$~~~~~~~~~$\lambda$~~~~~~~~~$\gamma$~~~~~~~~~$V_{0}$~~~~~~~~~$B_{0}$~~~~~~~~~$n$
~~~~~~~~~~~~~~~Positive roots ($X_{0}$)
\\
\hline\\
~~2~~~~~~~~~~0.1~~~~~~~~~0.1~~~~~~~~~1~~~~~~~~~~~~1~~~~~~~~~~~~1~~~~~~~~~~~~~~~~~~~~~~~~$-$
\\
~~,,~~~~~~~~~,,~~~~~~~~~,,~~~~~~~~,,~~~~~~~~~~~~,,~~~~~~~~~~~~~~2~~~~~~~~~~~~~~~~~~~~~~~~$-$
\\
~~,,~~~~~~~~~,,~~~~~~~~~,,~~~~~~~~,,~~~~~~~~~~~~,,~~~~~~~~~~~~~~4~~~~~~~~~~~~~~~~~~~~~~~~$-$
\\ \hline \\
~~1~~~~~~~~~~0.1~~~~~~~~~0.1~~~~~~~~~1~~~~~~~~~~~~1~~~~~~~~~~~~1~~~~~~~~~~~~~~~~~~~~~~~~$-$
\\
~~,,~~~~~~~~~,,~~~~~~~~~,,~~~~~~~~,,~~~~~~~~~~~~,,~~~~~~~~~~~~~~2~~~~~~~~~~~~~~~~~~~~~~~~$-$
\\
~~,,~~~~~~~~~,,~~~~~~~~~,,~~~~~~~~,,~~~~~~~~~~~~,,~~~~~~~~~~~~~~4~~~~~~~~~~~~~~~~~~~~~~~~$-$
\\ \hline \\
~~-2~~~~~~~~~~0.1~~~~~~~~~0.1~~~~~~~~~1~~~~~~~~~~~~1~~~~~~~~~~~~1~~~~~~~~~~~~~~~~~~~~~~~~$-$
\\
~~,,~~~~~~~~~,,~~~~~~~~~,,~~~~~~~~,,~~~~~~~~~~~~,,~~~~~~~~~~~~~~2~~~~~~~~~~~~~~~~~~~~~~~~
$0.978908$
\\
~~,,~~~~~~~~~,,~~~~~~~~~,,~~~~~~~~,,~~~~~~~~~~~~,,~~~~~~~~~~~~~~4~~~~~~~~~~~~~~~~~~~~~~~~
$0.792642$
\\ \hline \\
~~-3~~~~~~~~~~0.1~~~~~~~~~0.1~~~~~~~~~1~~~~~~~~~~~~1~~~~~~~~~~~~1~~~~~~~~~~~~~~~~~~~~~~~~
$1.7399$
\\
~~,,~~~~~~~~~,,~~~~~~~~~,,~~~~~~~~,,~~~~~~~~~~~~,,~~~~~~~~~~~~~~2~~~~~~~~~~~~~~~~~~~~~~~
~$1.5972$
\\
~~,,~~~~~~~~~,,~~~~~~~~~,,~~~~~~~~,,~~~~~~~~~~~~,,~~~~~~~~~~~~~~4~~~~~~~~~~~~~~~~~~~~~~~~
$1.3361$
\\ \hline \\
~~-4~~~~~~~~~~0.1~~~~~~~~~0.1~~~~~~~~~1~~~~~~~~~~~~1~~~~~~~~~~~~1~~~~~~~~~~~~~~~~~~~~~~~~
0.581281
\\
~~,,~~~~~~~~~,,~~~~~~~~~,,~~~~~~~~,,~~~~~~~~~~~~,,~~~~~~~~~~~~~~2~~~~~~~~~~~~~~~~~~~~~~~~
$0.4816$
\\
~~,,~~~~~~~~~,,~~~~~~~~~,,~~~~~~~~,,~~~~~~~~~~~~,,~~~~~~~~~~~~~~4~~~~~~~~~~~~~~~~~~~~~~~~
$0.4107$
\\
\hline\hline\\
\end{tabular}

\end{center}
\begin{center}
Table1c
\begin{tabular}{|l|}
\hline\hline ~~~~~~~~~~~~~~~~~~~~~~~~~~~~~~~~~~~~~~~~~~~~~~~~For $k=-0.5$ (dark energy)\\
\hline\hline
~~$\omega$~~~~~~~~~$\lambda$~~~~~~~~~$\gamma$~~~~~~~~~$V_{0}$~~~~~~~~~$B_{0}$~~~~~~~~~$n$
~~~~~~~~~~~~~~~Positive roots ($X_{0}$)
\\
\hline\\
~~2~~~~~~~~~~0.1~~~~~~~~~0.1~~~~~~~~~1~~~~~~~~~~~~1~~~~~~~~~~~~1~~~~~~~~~~~~~~~~~~~~~~~~$-$
\\
~~,,~~~~~~~~~,,~~~~~~~~~,,~~~~~~~~,,~~~~~~~~~~~~,,~~~~~~~~~~~~~~2~~~~~~~~~~~~~~~~~~~~~~~~$-$
\\
~~,,~~~~~~~~~,,~~~~~~~~~,,~~~~~~~~,,~~~~~~~~~~~~,,~~~~~~~~~~~~~~4~~~~~~~~~~~~~~~~~~~~~~~~$-$
\\ \hline \\
~~1~~~~~~~~~~0.1~~~~~~~~~0.1~~~~~~~~~1~~~~~~~~~~~~1~~~~~~~~~~~~1~~~~~~~~~~~~~~~~~~~~~~~~$-$
\\
~~,,~~~~~~~~~,,~~~~~~~~~,,~~~~~~~~,,~~~~~~~~~~~~,,~~~~~~~~~~~~~~2~~~~~~~~~~~~~~~~~~~~~~~~$-$
\\
~~,,~~~~~~~~~,,~~~~~~~~~,,~~~~~~~~,,~~~~~~~~~~~~,,~~~~~~~~~~~~~~4~~~~~~~~~~~~~~~~~~~~~~~~$-$
\\ \hline \\
~~-0.5~~~~~~~0.1~~~~~~~~~0.1~~~~~~~~~1~~~~~~~~~~~~1~~~~~~~~~~~~1~~~~~~~~~~~~~~~~~~~~~~~~$-$
\\
~~,,~~~~~~~~~,,~~~~~~~~~,,~~~~~~~~,,~~~~~~~~~~~~,,~~~~~~~~~~~~~~2~~~~~~~~~~~~~~~~~~~~~~~~$-$
\\
~~,,~~~~~~~~~,,~~~~~~~~~,,~~~~~~~~,,~~~~~~~~~~~~,,~~~~~~~~~~~~~~4~~~~~~~~~~~~~~~~~~~~~~~~$-$
\\ \hline \\
~~-2~~~~~~~~~~0.1~~~~~~~~~0.1~~~~~~~~~1~~~~~~~~~~~~1~~~~~~~~~~~~1~~~~~~~~~~~~~~~~~~~~~~~~1.0935
\\
~~,,~~~~~~~~~,,~~~~~~~~~,,~~~~~~~~,,~~~~~~~~~~~~,,~~~~~~~~~~~~~~2~~~~~~~~~~~~~~~~~~~~~~~~
$1.13196$
\\
~~,,~~~~~~~~~,,~~~~~~~~~,,~~~~~~~~,,~~~~~~~~~~~~,,~~~~~~~~~~~~~~4~~~~~~~~~~~~~~~~~~~~~~~~
$2.10243$
\\ \hline \\
~~-3~~~~~~~~~~0.1~~~~~~~~~0.1~~~~~~~~~1~~~~~~~~~~~~1~~~~~~~~~~~~1~~~~~~~~~~~~~~~~~~~~~~~~
1.00108
\\
~~,,~~~~~~~~~,,~~~~~~~~~,,~~~~~~~~,,~~~~~~~~~~~~,,~~~~~~~~~~~~~~2~~~~~~~~~~~~~~~~~~~~~~~~
$0.98201$
\\
~~,,~~~~~~~~~,,~~~~~~~~~,,~~~~~~~~,,~~~~~~~~~~~~,,~~~~~~~~~~~~~~4~~~~~~~~~~~~~~~~~~~~~~~~
$0.8391$
\\ \hline \\
~~-4~~~~~~~~~~0.1~~~~~~~~~0.1~~~~~~~~~1~~~~~~~~~~~~1~~~~~~~~~~~~1~~~~~~~~~~~~~~~~~~~~~~~~
0.963542
\\
~~,,~~~~~~~~~,,~~~~~~~~~,,~~~~~~~~,,~~~~~~~~~~~~,,~~~~~~~~~~~~~~2~~~~~~~~~~~~~~~~~~~~~~~~
$0.91852$
\\
~~,,~~~~~~~~~,,~~~~~~~~~,,~~~~~~~~,,~~~~~~~~~~~~,,~~~~~~~~~~~~~~4~~~~~~~~~~~~~~~~~~~~~~~
~$0.88349$
\\\hline\hline
 \end{tabular}

\end{center}
\begin{center}
Table1d
\begin{tabular}{|l|}

\hline\hline\\ ~~~~~~~~~~~~~~~~~~~~~~~~~~~~~~~~~~~~~~~~~~~~~~~~For $k=-1$ ($\Lambda$CDM)\\
\hline\hline
~~$\omega$~~~~~~~~~$\lambda$~~~~~~~~~$\gamma$~~~~~~~~~$V_{0}$~~~~~~~~~$B_{0}$~~~~~~~~~$n$
~~~~~~~~~~~~~~~Positive roots ($X_{0}$)
\\
\hline\\
~~2~~~~~~~~~~0.1~~~~~~~~~0.1~~~~~~~~~1~~~~~~~~~~~~1~~~~~~~~~~~~1~~~~~~~~~~~~~~~~~~~~~
~~~2.28511
\\
~~,,~~~~~~~~~,,~~~~~~~~~,,~~~~~~~~,,~~~~~~~~~~~~,,~~~~~~~~~~~~~~2~~~~~~~~~~~~~~~~~~~~~~~~
$0.214848$
\\
~~,,~~~~~~~~~,,~~~~~~~~~,,~~~~~~~~,,~~~~~~~~~~~~,,~~~~~~~~~~~~~~4~~~~~~~~~~~~~~~~~~~~~~~~
$0.214848$
\\ \hline \\
~~1~~~~~~~~~~0.1~~~~~~~~~0.1~~~~~~~~~1~~~~~~~~~~~~1~~~~~~~~~~~~1~~~~~~~~~~~~~~~~~~~~
~~~~2.67553
\\
~~,,~~~~~~~~~,,~~~~~~~~~,,~~~~~~~~,,~~~~~~~~~~~~,,~~~~~~~~~~~~~~2~~~~~~~~~~~~~~~~~~~~~~~~
$2.5398$
\\
~~,,~~~~~~~~~,,~~~~~~~~~,,~~~~~~~~,,~~~~~~~~~~~~,,~~~~~~~~~~~~~~4~~~~~~~~~~~~~~~~~~~~~~~
~$2.4112$
\\ \hline \\
~~-0.5~~~~~~~0.1~~~~~~~~~0.1~~~~~~~~~1~~~~~~~~~~~~1~~~~~~~~~~~~1~~~~~~~~~~~~~~~~~~~~~~~~$-$
\\
~~,,~~~~~~~~~,,~~~~~~~~~,,~~~~~~~~,,~~~~~~~~~~~~,,~~~~~~~~~~~~~~2~~~~~~~~~~~~~~~~~~~~~~~~$-$
\\
~~,,~~~~~~~~~,,~~~~~~~~~,,~~~~~~~~,,~~~~~~~~~~~~,,~~~~~~~~~~~~~~4~~~~~~~~~~~~~~~~~~~~~~~~$-$
\\ \hline \\
~~-2~~~~~~~~~~0.1~~~~~~~~~0.1~~~~~~~~~1~~~~~~~~~~~~1~~~~~~~~~~~~1~~~~~~~~~~~~~~~~~~~~~~~~1.01953
\\
~~,,~~~~~~~~~,,~~~~~~~~~,,~~~~~~~~,,~~~~~~~~~~~~,,~~~~~~~~~~~~~~2~~~~~~~~~~~~~~~~~~~~~~~~
$1.80231$
\\
~~,,~~~~~~~~~,,~~~~~~~~~,,~~~~~~~~,,~~~~~~~~~~~~,,~~~~~~~~~~~~~~4~~~~~~~~~~~~~~~~~~~~~~~~
$1.7093$
\\ \hline \\
~~-3~~~~~~~~~~0.1~~~~~~~~~0.1~~~~~~~~~1~~~~~~~~~~~~1~~~~~~~~~~~~1~~~~~~~~~~~~~~~~~~~~~~~~
0.947125
\\
~~,,~~~~~~~~~,,~~~~~~~~~,,~~~~~~~~,,~~~~~~~~~~~~,,~~~~~~~~~~~~~~2~~~~~~~~~~~~~~~~~~~~~~~~
$0.877062$
\\
~~,,~~~~~~~~~,,~~~~~~~~~,,~~~~~~~~,,~~~~~~~~~~~~,,~~~~~~~~~~~~~~4~~~~~~~~~~~~~~~~~~~~~~~~
$0.80149$
\\ \hline \\
~~-4~~~~~~~~~~0.1~~~~~~~~~0.1~~~~~~~~~1~~~~~~~~~~~~1~~~~~~~~~~~~1~~~~~~~~~~~~~~~~~~~~~~~~
0.930215
\\
~~,,~~~~~~~~~,,~~~~~~~~~,,~~~~~~~~,,~~~~~~~~~~~~,,~~~~~~~~~~~~~~2~~~~~~~~~~~~~~~~~~~~~~~~
$0.739$
\\
~~,,~~~~~~~~~,,~~~~~~~~~,,~~~~~~~~,,~~~~~~~~~~~~,,~~~~~~~~~~~~~~4~~~~~~~~~~~~~~~~~~~~~~~~
$0.6283$
\\\hline\hline

\end{tabular}

\end{center}
\begin{center}
Table1e
\begin{tabular}{|l|}
\hline\hline\\
 ~~~~~~~~~~~~~~~~~~~~~~~~~~~~~~~~~~~~~~~~~~~~~~~~For $k=-2$ (phantom)\\
\hline\hline
~~$\omega$~~~~~~~~~$\lambda$~~~~~~~~~$\gamma$~~~~~~~~~$V_{0}$~~~~~~~~~$B_{0}$~~~~~~~~~$n$
~~~~~~~~~~~~~~~Positive roots ($X_{0}$)
\\
\hline\\
~~2~~~~~~~~~~0.1~~~~~~~~~0.1~~~~~~~~~1~~~~~~~~~~~~1~~~~~~~~~~~~1~~~~~~~~~~~~~~~~~~~~~~~
~1.99826
\\
~~,,~~~~~~~~~,,~~~~~~~~~,,~~~~~~~~,,~~~~~~~~~~~~,,~~~~~~~~~~~~~~2~~~~~~~~~~~~~~~~~~~~~~~
~$1.7602$
\\
~~,,~~~~~~~~~,,~~~~~~~~~,,~~~~~~~~,,~~~~~~~~~~~~,,~~~~~~~~~~~~~~4~~~~~~~~~~~~~~~~~~~~~~~
~$1.5221$
\\ \hline \\
~~1~~~~~~~~~~0.1~~~~~~~~~0.1~~~~~~~~~1~~~~~~~~~~~~1~~~~~~~~~~~~1~~~~~~~~~~~~~~~~~~~~~~~
~1.93983
\\
~~,,~~~~~~~~~,,~~~~~~~~~,,~~~~~~~~,,~~~~~~~~~~~~,,~~~~~~~~~~~~~~2~~~~~~~~~~~~~~~~~~~~~~~
~$1.72951$
\\
~~,,~~~~~~~~~,,~~~~~~~~~,,~~~~~~~~,,~~~~~~~~~~~~,,~~~~~~~~~~~~~~4~~~~~~~~~~~~~~~~~~~~~~~
~$1.5291$
\\ \hline \\
~~-0.5~~~~~~~0.1~~~~~~~~~0.1~~~~~~~~~1~~~~~~~~~~~~1~~~~~~~~~~~~1~~~~~~~~~~~~~~~~~~~~~~~~2.77177
\\
~~,,~~~~~~~~~,,~~~~~~~~~,,~~~~~~~~,,~~~~~~~~~~~~,,~~~~~~~~~~~~~~2~~~~~~~~~~~~~~~~~~~~~~~
~$1.4193$
\\
~~,,~~~~~~~~~,,~~~~~~~~~,,~~~~~~~~,,~~~~~~~~~~~~,,~~~~~~~~~~~~~~4~~~~~~~~~~~~~~~~~~~~~~~
~$1.3629$
\\ \hline \\
~~-2~~~~~~~~~~0.1~~~~~~~~~0.1~~~~~~~~~1~~~~~~~~~~~~1~~~~~~~~~~~~1~~~~~~~~~~~~~~~~~~~~~~~
~0.937459
\\
~~,,~~~~~~~~~,,~~~~~~~~~,,~~~~~~~~,,~~~~~~~~~~~~,,~~~~~~~~~~~~~~2~~~~~~~~~~~~~~~~~~~~~~
~~$0.88362$
\\
~~,,~~~~~~~~~,,~~~~~~~~~,,~~~~~~~~,,~~~~~~~~~~~~,,~~~~~~~~~~~~~~4~~~~~~~~~~~~~~~~~~~~~~~~
$0.76301$
\\ \hline \\
~~-3~~~~~~~~~~0.1~~~~~~~~~0.1~~~~~~~~~1~~~~~~~~~~~~1~~~~~~~~~~~~1~~~~~~~~~~~~~~~~~~~~~~~~
0.916792
\\
~~,,~~~~~~~~~,,~~~~~~~~~,,~~~~~~~~,,~~~~~~~~~~~~,,~~~~~~~~~~~~~~2~~~~~~~~~~~~~~~~~~~~~~
~~$0.83928$
\\
~~,,~~~~~~~~~,,~~~~~~~~~,,~~~~~~~~,,~~~~~~~~~~~~,,~~~~~~~~~~~~~~4~~~~~~~~~~~~~~~~~~~~~~~~
$0.800702$
\\ \hline \\
~~-4~~~~~~~~~~0.1~~~~~~~~~0.1~~~~~~~~~1~~~~~~~~~~~~1~~~~~~~~~~~~1~~~~~~~~~~~~~~~~~~~~~~~
~0.921105
\\
~~,,~~~~~~~~~,,~~~~~~~~~,,~~~~~~~~,,~~~~~~~~~~~~,,~~~~~~~~~~~~~~2~~~~~~~~~~~~~~~~~~~~~~~
~$0.901903$
\\
~~,,~~~~~~~~~,,~~~~~~~~~,,~~~~~~~~,,~~~~~~~~~~~~,,~~~~~~~~~~~~~~4~~~~~~~~~~~~~~~~~~~~~~~~
$0.8741$
\\
 \hline\hline
\end{tabular}

\end{center}
~~~~~~~~~~~~~~~{\bf Table 1a-e:} Values of $X_{0}$ for different
values of parameters
$\lambda,\gamma,V_{0},B_{0},n,\omega,k$.\\
\vspace{.5cm}

\begin{figure}
~~~~~~~~~~~~~~~~~~~~~~Fig.1a~~~~~~~~~~~~~~~~~~~~~~~~~~~~~~~~~~~~~~~
Fig.1b~~~~~~~~~~~~~~~~~~~~~~~~~~~~~~~~~~~~~~~Fig.1c\\
\includegraphics[height=2in, width=2in]{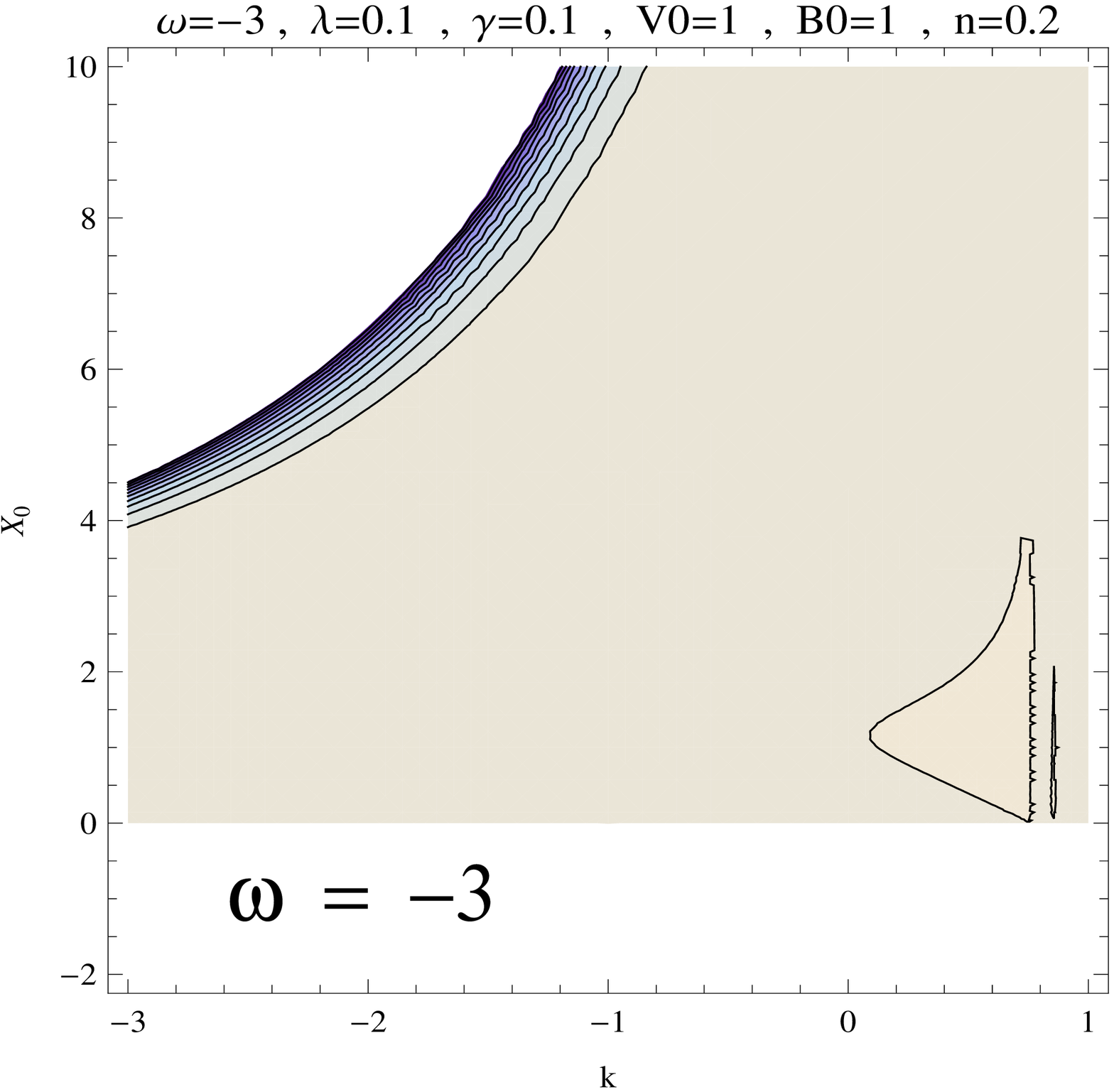}~~~~~~~
\includegraphics[height=2in, width=2in]{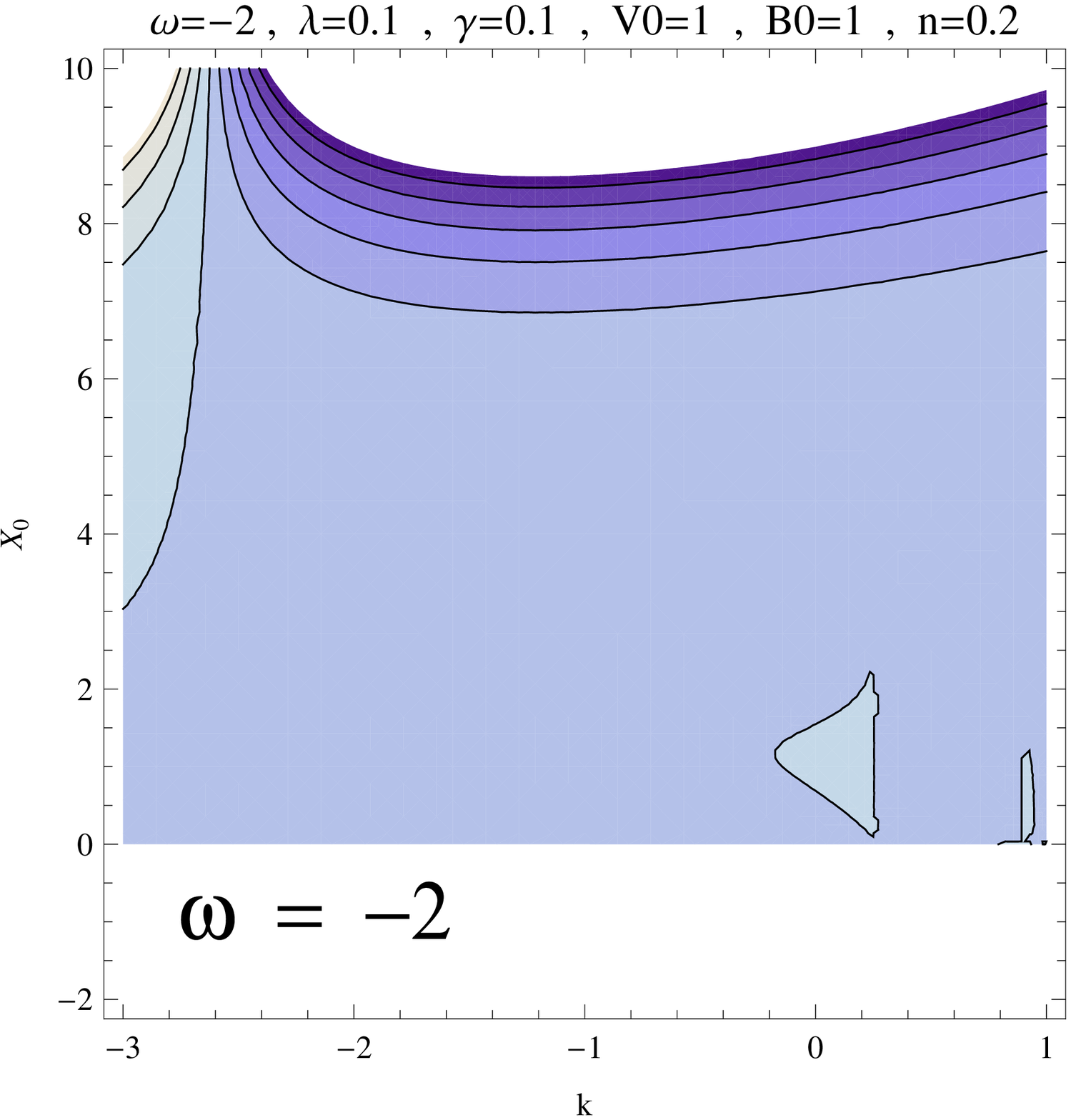}~~~~~~~
\includegraphics[height=2in, width=2in]{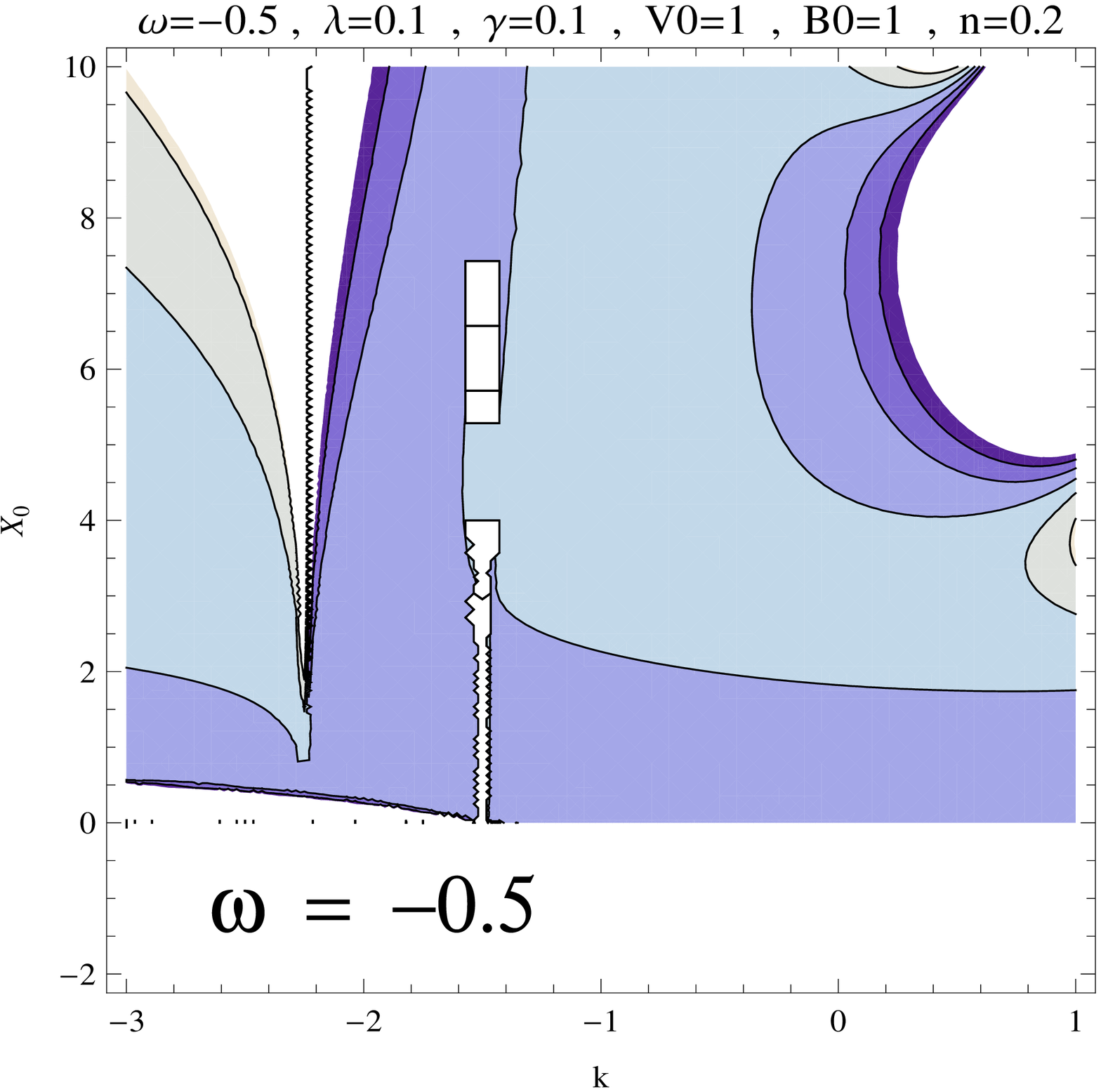}\\
~~~~~~~~~~~~~~~~~~~~~~Fig.1d~~~~~~~~~~~~~~~~~~~~~~~~~~~~~~~~~~~~~~~
Fig.1e~~~~~~~~~~~~~~~~~~~~~~~~~~~~~~~~~~~~~~~Fig.1f\\
\includegraphics[height=2in, width=2in]{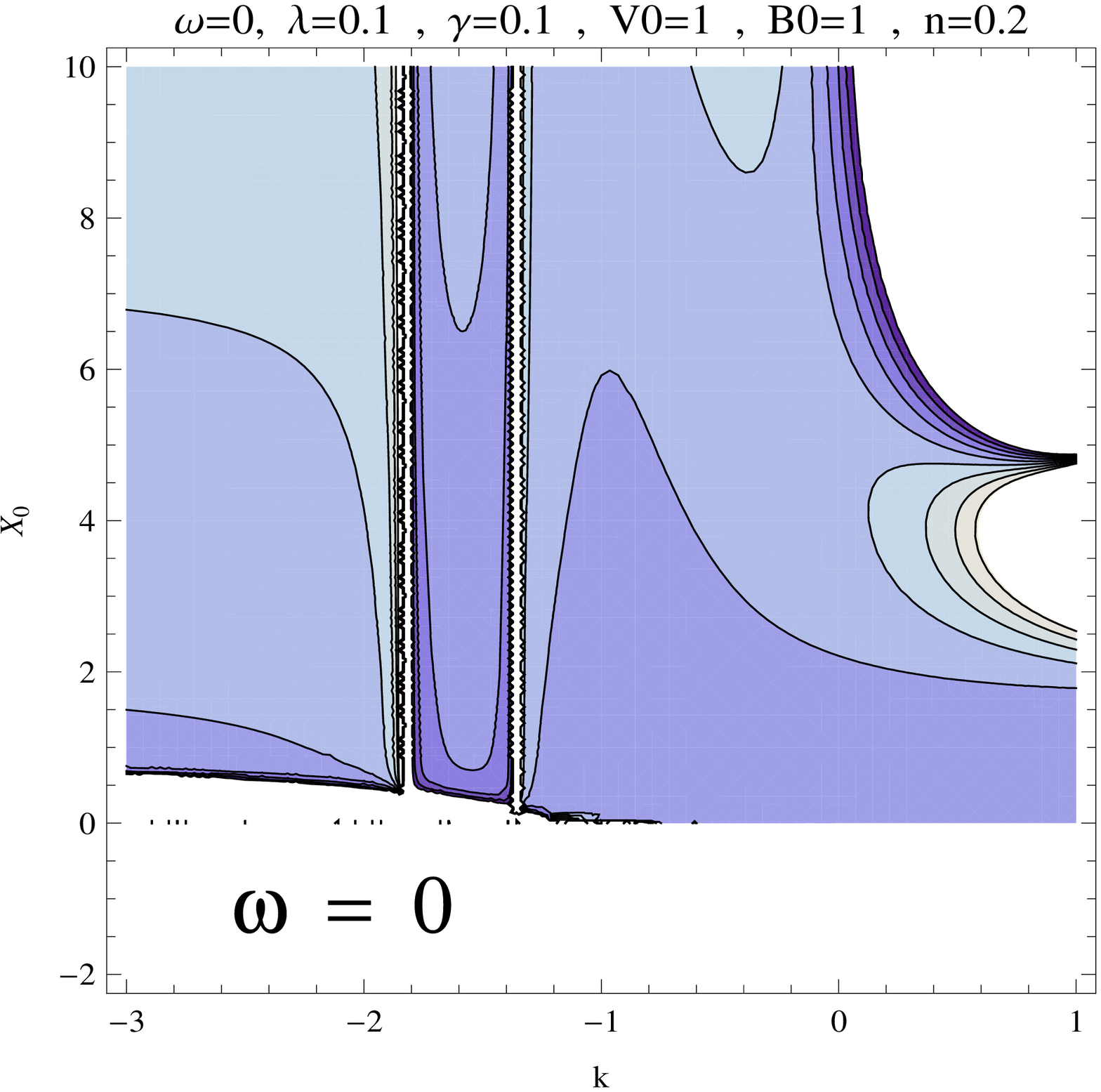}~~~~~~~
\includegraphics[height=2in, width=2in]{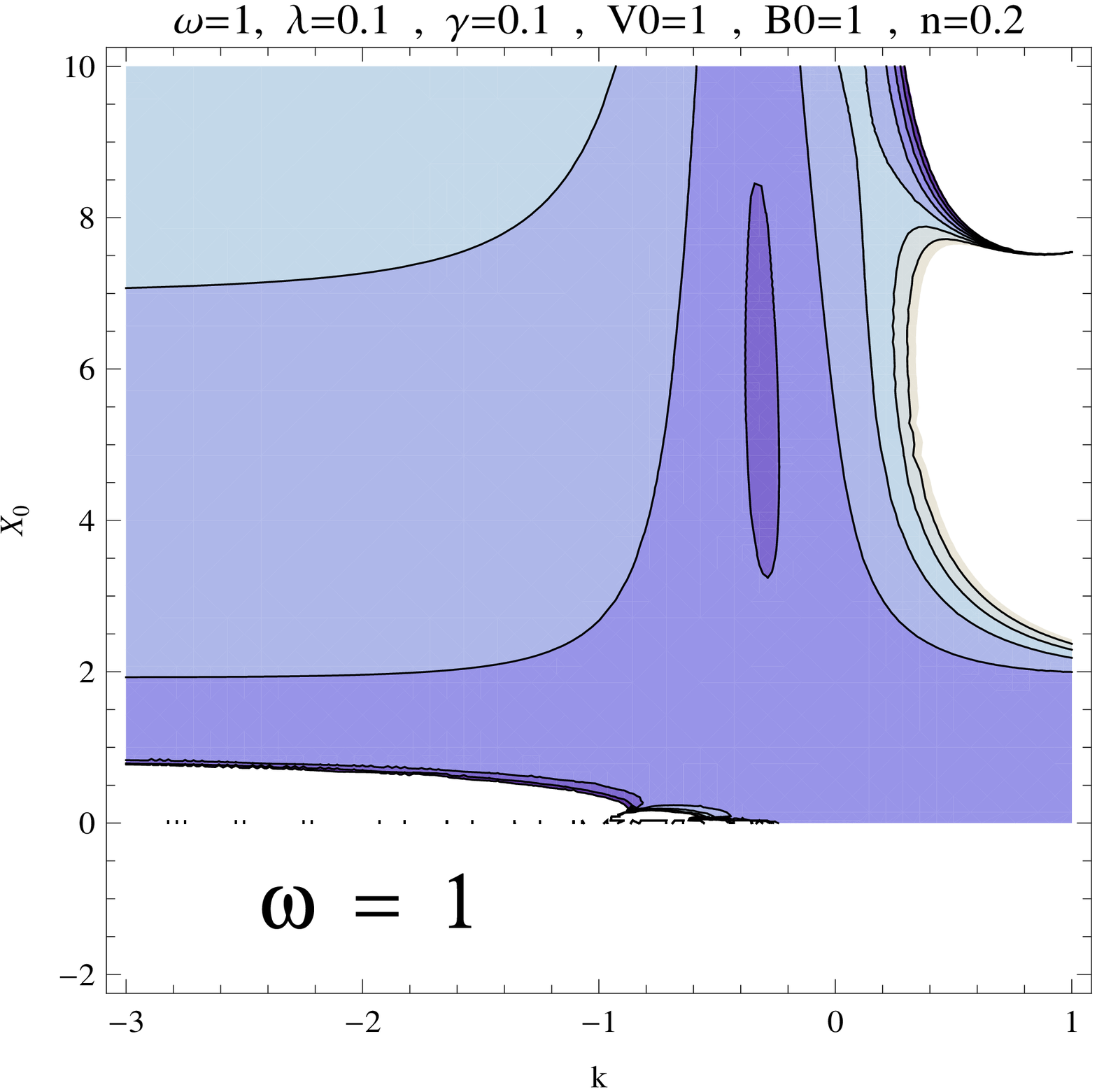}~~~~~~~
\includegraphics[height=2in, width=2in]{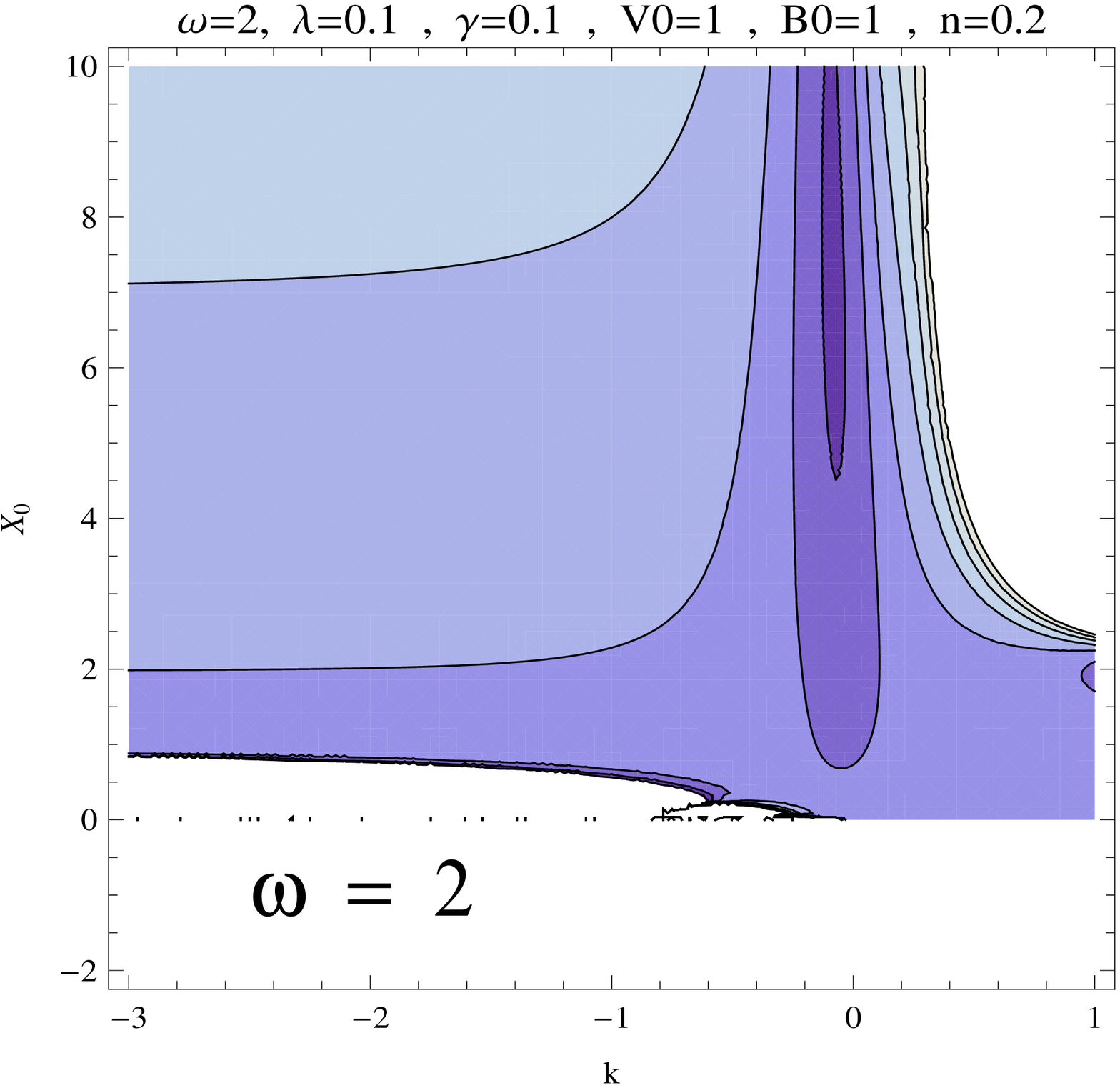}\\
~~~~~~~~~~~~~~~~~~~~~~Fig.1g~~~~~~~~~~~~~~~~~~~~~~~~~~~~~~~~~~~~~~~
Fig.1h~~~~~~~~~~~~~~~~~~~~~~~~~~~~~~~~~~~~~~~Fig.1i\\
\includegraphics[height=2in, width=2in]{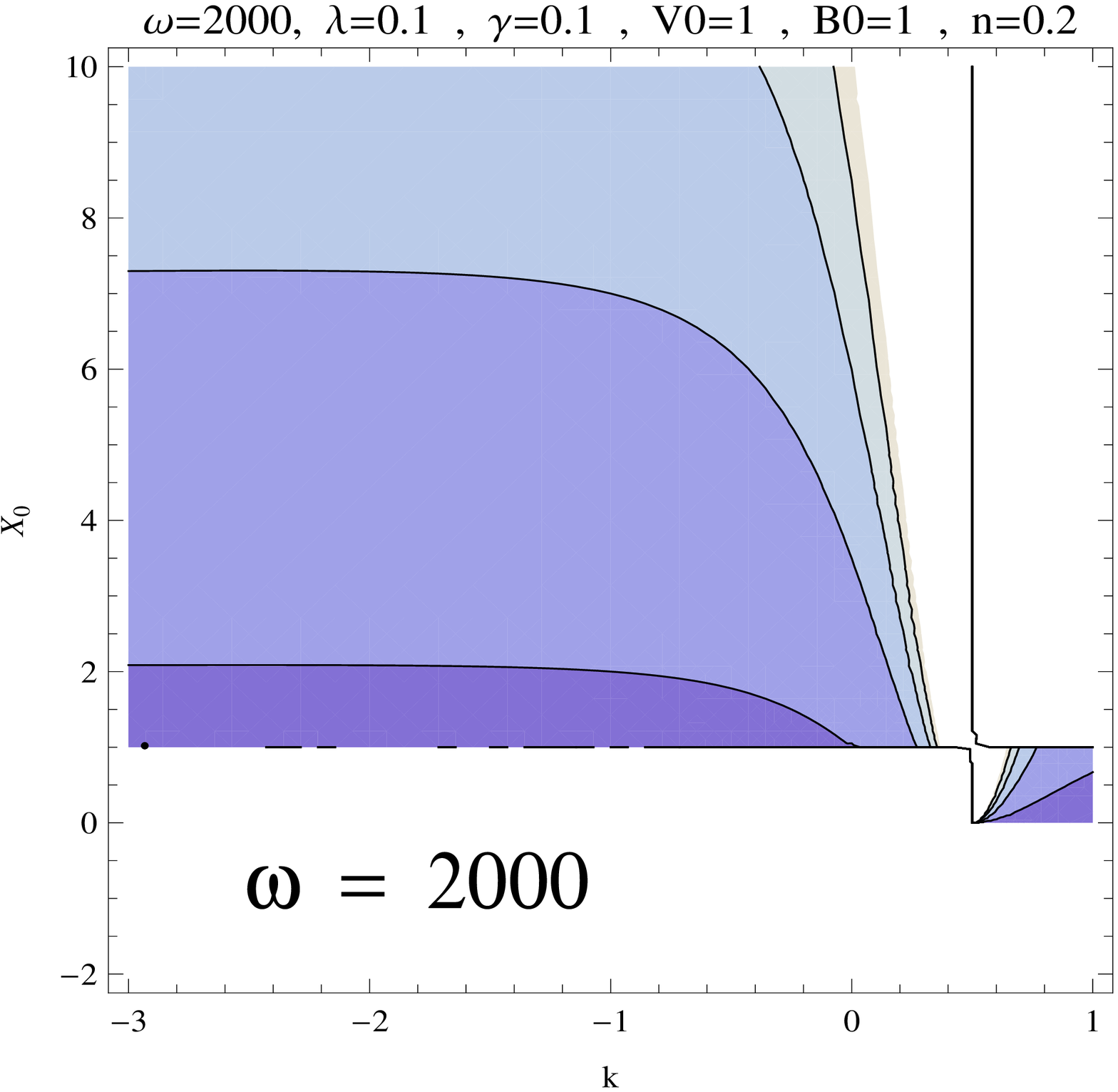}~~~~~~~
\includegraphics[height=2in, width=2in]{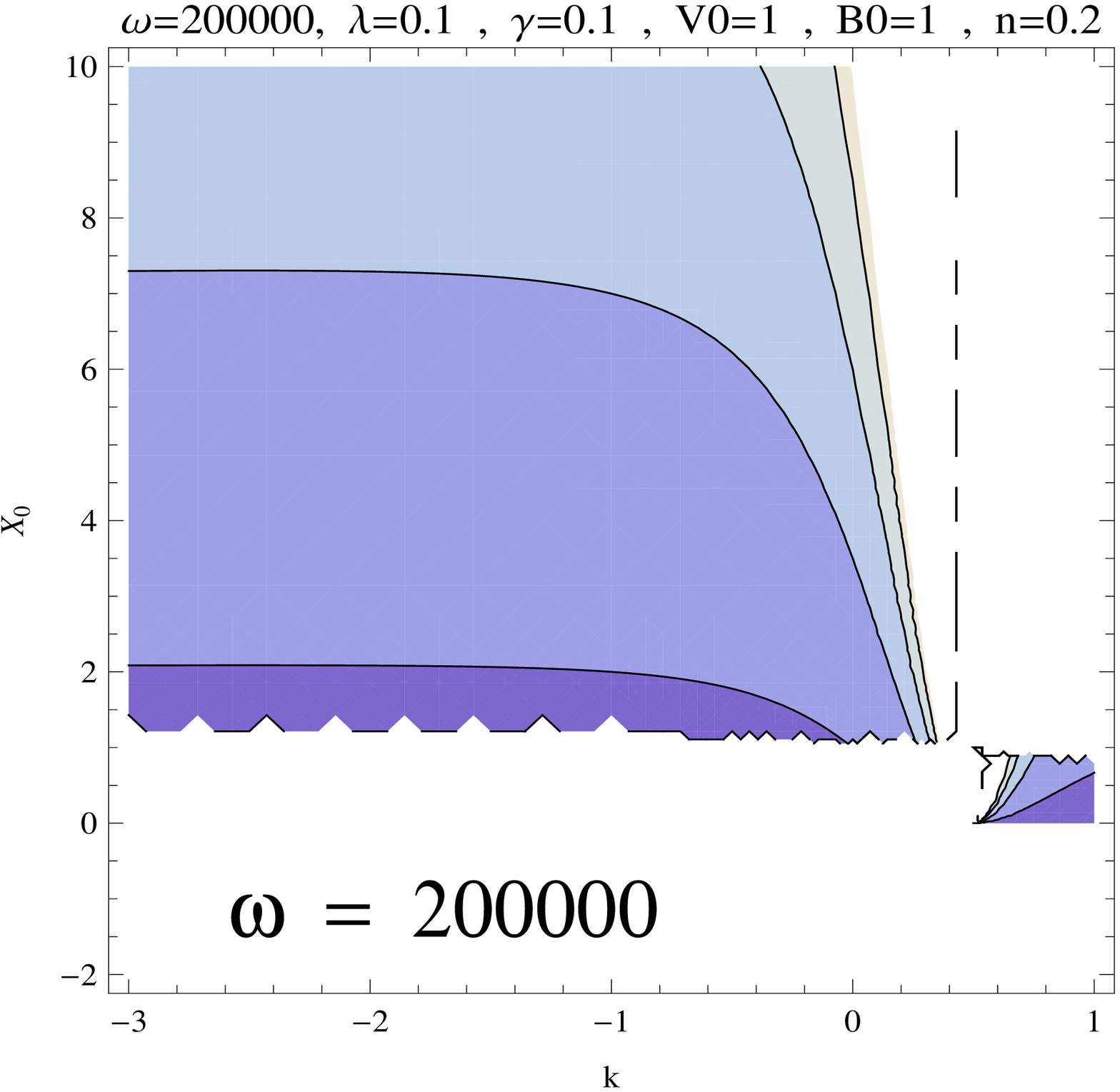}~~~~~~~
\includegraphics[height=2in, width=2in]{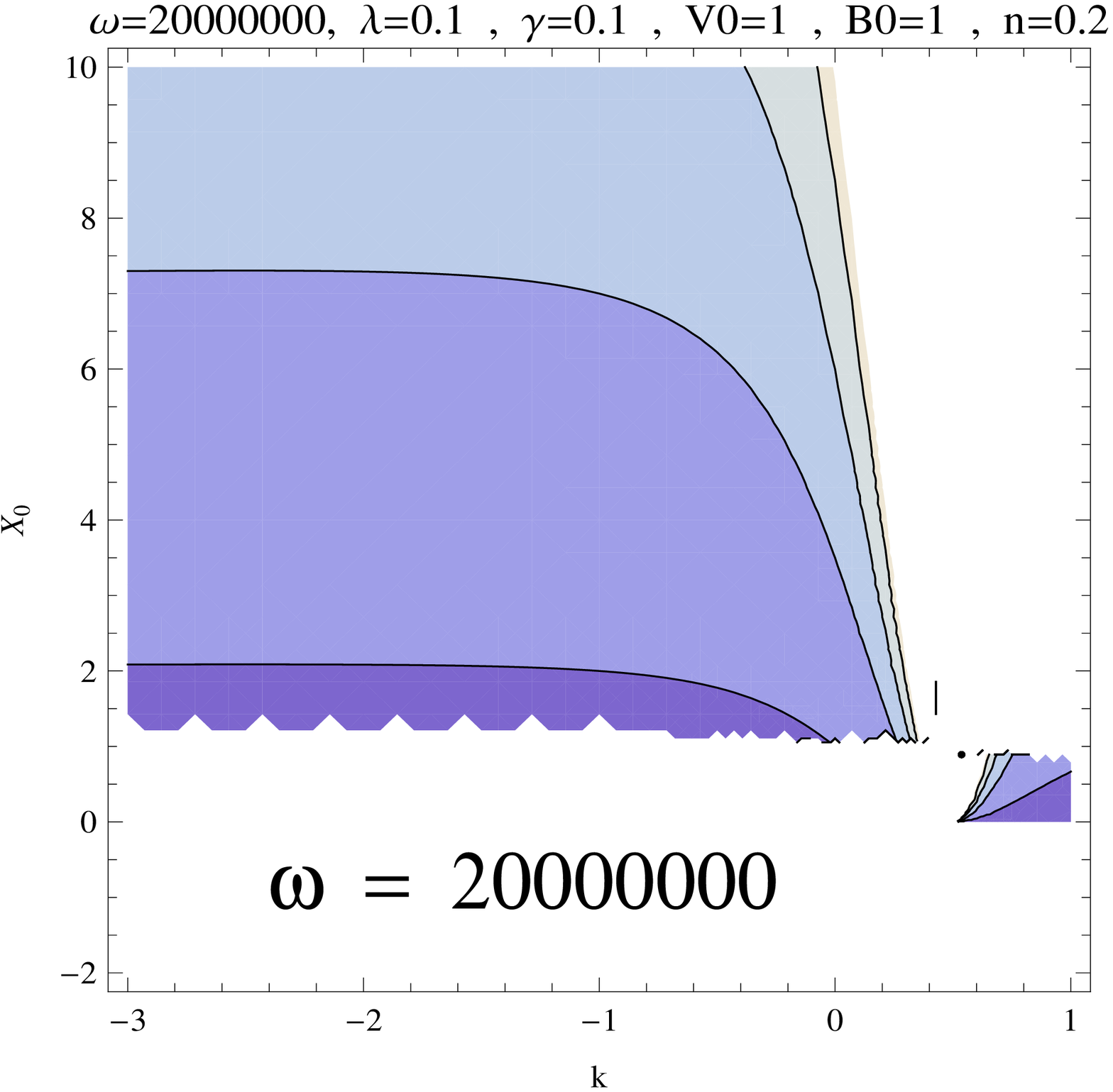}\\

\vspace{1cm} {\bf Figs. 1a-i} show the $X_0$ contours in the
$k$-$X_0$ plane for different values of $\omega$.   \vspace{2cm}
\end{figure}

Here, the two very important parameters are $\omega$, the
Brans-Dicke parameter and the EoS parameter, $k$. With a slight
variation of all other parameters we have considered two extremal
variations for $\omega$ and $k$. For EoS parameter $k$ being
positive ($=1$) one we get no positive solution irrespective of
the value of the $\omega$ in the range $-3<\omega<2$ if $n$ is
considered to be unity. But for higher values of $n$ there is a
possibility of having positive solutions for highly negative
$\omega$ cases, i.e., for greater deviations from Einstein
gravity. But highly negative values of $\omega$ are not physical
either, since they produce ghost. In radiation era
($k=\frac{1}{3}$), for highly negative values of $\omega$ positive
roots are found. However, $\omega>-2$ contributes no positive
solutions in the radiation era. In expanding universe the same
trend continues except the fact that here the lower limit of the
range of $\omega$ that gives non positive solution rises compared
to the radiation era. At phantom crossing and phantom era we
always have positive solutions (for the range $-4<\omega$).
Physically interpreting we can say when positive pressure fluids
are there in the universe the outcome of a collapse is more likely
to be a NS. Now if we lower the value of the Brans Dicke
parameter, i.e., we go far from the Einstein gravity the chances
of having a NS increases even for a non-expanding universe. In
fact negative values of $\omega$ almost confirms the possibility
of having a NS. Leaving the positive pressure zone behind as we
look into the quintessence/ phantom era i.e. we consider the
expanding universe (rather to say the scenario of cosmic
acceleration) it immediately gives NS as the only final fate of
the collapse. Here, even though we increase $\omega$ to a notable
higher value (i.e., we move towards the Einstein gravity) yet the
collapse results in NS only. As only two parameters are
controlling the end fate, we will plot their variations in the fig
$1a-i$. In figures 1a to 1i we have plotted the $k-X_0$ contours
for increasing values of $\omega$. $0<k<1$ zone is less probable
zone to have a contour for negative values of $\omega$. But with
positive and higher values of $\omega$ contours are there over the
whole range of $k$. Now whatever be the value of $\omega$ we get
contours for negative $k$-s.

\section{Conclusion}\label{Discussion}
In this work, we have assumed the spherically symmetric space time
model with Vaidya null radiation and perfect fluid. We have
determined the solution of Einstein equation in Brans-Dicke
gravity with self interacting potential and after choosing power
law form of the potential we have determined the Husain metric
(generalized Vaidya metric). Next step was to inspect the
existence of the radial null geodesic from the final fate of the
collapsing object. Existence of such geodesic points corresponds
to the chances of having a NS. If not, then the possibility of a
BH is confirmed. In this study the general trend is to have a NS
if we consider late time universe and smaller values of
Brans-Dicke coupling constant. Higher positive values of $\omega$
correspond to lesser deviation from Einstein gravity, whereas the
lower negative values points towards extreme deviations and
greater modification of the gravity theory. So when we decrease
$\omega$ the probability of having a NS, as an end state of
collapse becomes greater, even if a non-accelerating scenario is
considered. It should be realized that sufficiently small
$|\omega|$ turns the Brans-Dicke field sufficiently dynamic giving
such an outcome. The trend of collapse for late time non-positive
values of $k$ matches with the work of Scheel (Scheel, M. A. et
al. 1995). {\bf The potential working around a compact object
increases the speed of the flow around it and gradually makes it
supersonic via sonic point crossing. Particularly the
pseudo-Newtonian potential working around the object makes the
velocity of the flow equal to the speed of light near the event
horizon in case of a black hole since it does not have any hard
surface like neutron stars. But in case of a NS the flow will not
get absorbed at any horizon and near the singularity the flow will
be exotic showing abrupt/non-uniform spectra. Such sources may be
found in the cases of AGNs. Kovacs, Z., Harko, T. (2010) have
proposed such a phenomenon. A similar topic was discussed by
Virbhadra, K. S., Keeton, C. R. (2008) where they concluded that
the lensing characteristics of strongly naked singularities are
qualitatively very different from those due to Schwarzschild black
holes. In this context, it is worth mentioning that if it is
possible to test or compare the model described in the present
assignment with the methods described in the above mentioned
references, then it will be of interest to discuss their origin
and may be our current results will get a stronger base and some
astrophysical support. For the time being we keep it an open
question worthy of addressing in near future.}

\begin{contribution}\\\\
{\bf Acknowledgement :}\\

Authors thank IUCAA, Pune for local hospitality and research
facilities, where a part of this work was done. RB thanks CSIR for
awarding Research Associate fellowship and UD thanks CSIR project
``Dark Energy Models and Accelerating Universe" $(No.~
03(1206)/12/EMR-II)$. The authors also thank the anonymous referee
for his or her constructive comments that helped them to improve
the quality of the manuscript.

\end{contribution}

\section{Appendix}

\textbf{Equation (18) is a second order Non-homogeneous Ordinary
differential equation with variable coefficient (r). If we
multiply eqn. (18) by $r^{3}$, we see that it reduces to
Cauchy-Euler differential equation. So we make the substitution,
\begin{equation}
r=e^{z}
\end{equation}
This transforms the equation into a second order differential
equation with constant coefficients, which of course can be solved
by the known methods. We take the corresponding homogeneous
differential equation and find the Complementary function (C.F.)
as given below,
\begin{equation}
C.F. (m_{c})=f_{1}(t)r^{\omega_{1}}+f_{2}(t)r^{\omega_{2}}
\end{equation}
where
\begin{equation}
\omega_{1},
\omega_{2}=\frac{\left(\omega-3k-2k\omega-1\right)\pm\sqrt{\left\{\left(k+2\right)+\left(2k-1\right)\left(\omega+1\right)\right\}^{2}-4\left(6k\omega+5k+9\omega+8\right)}}{2\left(\omega+1\right)}
\end{equation}
Here $f_{1}(t)$ and $f_{2}(t)$ are arbitrary functions of $t$ as
stated earlier. Then we go on to calculate the Particular Integral
(P.I.) for the non-homogeneous part of the differential equation
as given below,
\begin{equation}
P.I. (m_{p})=\frac{5\omega
k+4k+7\omega+6}{\left(1-\omega_{1}\right)\left(1-\omega_{2}\right)}r-\frac{2\dot{B}\left(1+k\right)\left(1+\omega\right)}
{B\left(2-\omega_{1}\right)\left(2-\omega_{2}\right)}r^{2}-\frac{V_{0}B^{n-1}\left(\omega+1\right)^{4}
\left(k+1\right)}{\left\{\left(\omega+n\right)-\omega_{1}\left(\omega+1\right)\right\}\left\{\left(\omega+n\right)
-\omega_{2}\left(\omega+1\right)\right\}}r^{\frac{\omega+n}{\omega+1}}
\end{equation}
Finally we add the C.F. $(m_{c})$ and P.I. $(m_{p})$ to get the
final expression for $m$ in eqn. 19, i.e. $m=m_{c}+m_{p}$.}

\end{document}